Electron acceleration to relativistic energies at a strong quasi-parallel shock wave

A. Masters,[1]* L. Stawarz,[1] M. Fujimoto,[1] S. J. Schwartz,[2] N. Sergis,[3]

M. F. Thomsen,[4] A. Retinò,[5] H. Hasegawa,[1] G. R. Lewis,[6,7] A. J. Coates,[6,7] P. Canu,[5]

M. K. Dougherty[2]

[1]Institute of Space and Astronautical Science, Japan Aerospace Exploration Agency,

3-1-1 Yoshinodai, Chuo-ku, Sagamihara, Kanagawa 252-5210, Japan.

[2]Space and Atmospheric Physics Group, The Blackett Laboratory, Imperial College London,

Prince Consort Road, London, SW7 2AZ, UK.

[3]Office of Space Research and Technology, Academy of Athens, Soranou Efesiou 4, 11527

Athens, Greece.

[4]Space Science and Applications, Los Alamos National Laboratory, Los Alamos, New

Mexico, 87545, USA.

[5]Laboratoire de Physique des Plasmas, Centre National de la Recherche Scientifique,

Observatoire de Saint-Maur, 4 avenue de Neptune, Saint-Maur-Des-Fossés, 94107 France.

[6]Mullard Space Science Laboratory, Department of Space and Climate Physics, University

College London, Holmbury St. Mary, Dorking, Surrey, RH5 6NT, UK.

[7]The Centre for Planetary Sciences at UCL/Birkbeck, Gower St., London, WC1E 6BT, UK.

*Please address correspondence to A.M.

a.masters@stp.isas.jaxa.jp

Introductory paragraph:

Electrons can be accelerated to ultrarelativistic energies at strong (high-Mach number) collisionless shock waves that form when stellar debris rapidly expands after a supernova[1-3]. Collisionless shock waves also form in the flow of particles from the Sun (the solar wind), and extensive spacecraft observations have established that electron acceleration at these shocks is effectively absent whenever the upstream magnetic field is roughly parallel to the shock surface normal (quasi-parallel conditions)[4-8]. However, it is unclear whether this magnetic dependence of electron acceleration also applies to the far stronger shocks around young supernova remnants, where local magnetic conditions are poorly understood. Here we present Cassini spacecraft observations of an unusually strong solar system shock wave (Saturn's bow shock) where significant local electron acceleration has been confirmed under quasi-parallel magnetic conditions for the first time, contradicting the established magnetic dependence of electron acceleration at solar system shocks[4-8]. Furthermore, the acceleration led to electrons at relativistic energies (~MeV), comparable to the highest energies ever attributed to shock-acceleration in the solar wind[4]. These observations demonstrate that at high-Mach numbers, like those of young supernova remnant shocks, quasi-parallel shocks become considerably more effective electron accelerators.

Text:

Shock waves form when flow relative to an obstacle is greater than the speed of information transfer via the medium. Flow kinetic energy is dissipated at a shock, and fluid properties change abruptly, producing a slower downstream flow that is able to avoid the obstacle. In collisional fluids (like Earth's atmosphere) energy dissipation at a shock wave occurs via inter-particle collisions; however, in effectively collisionless (highly tenuous) media, such as charged particle (plasma) space environments, dissipation at shocks is via particle-electromagnetic field interactions[9]. The fraction of flow kinetic energy dissipated at a shock is indicated by the shock Mach numbers (flow speed divided by upstream wave speeds) and particle motion at the shock is controlled by the shock angle ($\theta_{Bn}$), the angle between the shock normal and the upstream magnetic field vector, which defines two categories of collisionless shock: quasi-parallel ($0° < \theta_{Bn} < 45°$) and quasi-perpendicular ($45° < \theta_{Bn} < 90°$). Energy dissipation at a collisionless shock not only leads to heating of the bulk plasma, but can also accelerate some particles to much higher energies.

It is widely believed that a major fraction of the energetic charged particles that pervade the Galaxy ("cosmic rays" with energies up to ~$10^{15}$ eV) are accelerated at collisionless shock waves associated with supernova explosions[1]. The specific acceleration mechanism typically invoked in this context, commonly referred to as diffusive shock acceleration, is thought to result from a Fermi process where particles bounce between converging scattering centres (e.g. electromagnetic waves) located on either side of the shock[1]. Emissions associated with ultrarelativistic electrons produced at young (<1000 year-old) supernova remnant shocks have been comprehensively studied using both Earth-based and space-based telescopes[2,3,10]. However, poorly constrained local conditions at these exotic, distant shocks[11], and in particular the hardly known magnetic field conditions lead to uncertainty surrounding the electron acceleration process.

Observations made by spacecraft during encounters with collisionless shocks in the Solar System can potentially shed light on the physics of these supernova remnant shocks[9]. Shocks are common in the solar wind plasma that flows away from the Sun and carries the solar magnetic field into interplanetary space[12,13]. Electron acceleration is often observed, although not to the ultrarelativistic (TeV-PeV) energies produced at young supernova remnant shocks, and various acceleration mechanisms have been discussed. Observed electron acceleration is significantly more efficient at quasi-perpendicular shocks than at quasi-parallel shocks[4-8], although even under quasi-perpendicular conditions the detection of relativistic (MeV) electrons is rare[4]. Since solar system shocks are generally far weaker (lower Mach number) than their young supernova remnant counterparts, and also far smaller, it is unclear what the observed magnetic dependence of electron acceleration means for much stronger shocks. There is some limited indication, and it has been predicted, that quasi-parallel shocks may become efficient electron accelerators at high-Mach numbers[5,14,15].

Here we present observations of an unusually strong quasi-parallel shock wave made by NASA's Cassini spacecraft. Cassini has made hundreds of crossings of the shock wave that stands in the solar wind in front of Saturn (the planetary bow shock) due to the obstacle presented by the planet's intrinsic magnetic field (the planetary magnetosphere)[16-18]. Fig. 1 illustrates where the spacecraft was located at ~01:10 Universal Time (UT) on 3 February 2007 when it encountered Saturn's bow shock under atypical conditions. In situ observations made by Cassini between 00:00 and 02:00 UT are shown in Fig. 2. The spacecraft began this interval upstream of Saturn's bow shock and ended downstream, making an inbound shock crossing (illustrated in Fig. 1). Because the purpose of this letter is to present evidence for electron acceleration at the shock, non-electron data are used as (necessary) diagnostics. Hereafter, "low-energy" electrons refer to electrons at energies below 18 keV, and "high-energy" electrons are above 18 keV, based on the energy ranges of the two electron detectors (Fig. 2d and 2e).

The magnetic field strength upstream of this shock encounter was ~0.1 nT (1 μG). The determination of further shock parameters from Cassini observations and a solar wind model has been previously reported to give an Alfvén Mach number ($M_A$, related to the upstream speed of Alfvén waves, one of the fundamental wave modes in a magnetized plasma) of ~100, which largely results from the weak upstream magnetic field strength[18]. This $M_A$ value is very high for a solar system shock, and approaches the higher Mach number regime of young supernova remnant shocks. Using typical electron and ion temperatures in the near-Saturn solar wind[16] we obtain ~25 for both the sonic and fast magnetosonic Mach numbers, which are also relatively high for a solar system shock.

Before ~00:05 UT the spacecraft was not magnetically connected to Saturn's bow shock, and the magnetic field orientation suggests that the shock (located planetward of the spacecraft) was quasi-perpendicular at this time ($\theta_{Bn}$~60°). At ~00:05 UT the field orientation changed, magnetically connecting the spacecraft to the bow shock and making the shock quasi-parallel ($\theta_{Bn}$~20°). The subsequent shock transition signature is largely typical of quasi-parallel shocks[19,20]. The motion of shock-reflected ions back upstream lead to an interaction with the incoming solar wind ions, resulting in strong upstream (before ~01:10 UT) magnetic field fluctuations (Fig. 2a) and a population of so-called "diffuse" ions (Fig. 2g, detected even though ion detector pointing was not appropriate to also detect the incoming ion flow). Surrounding ~01:10 UT there was an extended magnetic transition into a stronger downstream (after ~01:10 UT) field environment (Fig. 2a), where there is clear evidence for the presence of shock-heated and compressed bulk solar wind plasma (Fig. 2e and 2g).

Cassini intermittently observed emissions between the electron gyrofrequency and electron plasma frequency during the interval (Fig. 2c). Firm conclusions about directional particle fluxes or pitch angle distributions cannot be drawn due to instrument pointing limitations and the variability of the magnetic field orientation.

Fig. 3 compares average electron energy spectra measured during different two-minute-long intervals, and Fig. 4 compares the spectra measured by the high-energy electron detector only. A high-energy electron population (which gave above-background intensities up to ~100 keV) was detected from when the spacecraft became magnetically connected to the shock at ~00:05 UT (Fig. 2d and 3a). This population appears to have been measured both upstream and downstream of the shock (Fig. 3a, 3c, and 4), and has a similar spectral shape to the (higher intensity) spectrum measured immediately inside Saturn's magnetosphere less than an hour earlier (Fig. 4). We identify this high-energy electron population as predominantly electrons escaping from Saturn's magnetosphere, where they were already at similar energies[21]. The observation of magnetospheric particle escape requires a magnetic connection to the bow shock[21], consistent with the detection of this population from ~00:05 UT onwards.

Immediately after the major low-energy electron heating had occurred (at ~01:10 UT) the low-energy electron detector measured intensities clearly above the one-count level between 5 and 10 keV (Fig. 2f and 3b). This signature of energisation of low-energy solar wind electrons near the shock transition is a well-observed phenomenon at (weaker) quasi-perpendicular shocks[5,8], and its presence at this (stronger) quasi-parallel shock is consistent with predictions about this higher $M_A$ regime[5,14,15]. The (more sensitive) high-energy electron detector observed a broad signature centred on the time of low-energy electron energisation (Fig. 2d and 2f). This high-energy signature was at least 10 times more intense than that of the escaping magnetospheric electrons in all relevant energy channels (<100 keV), and produced above-background intensities in the higher energy (>100 keV) channels, the highest of which approaches MeV energies (Fig. 4). We identify this combined electron signature as the first in situ evidence for acceleration of solar wind electrons to relativistic energies at a quasi-parallel solar system shock. Although escaping magnetospheric electrons were likely also accelerated to >100 keV, the far greater intensity of the signal of shock-accelerated solar wind electrons at <100 keV in the high-energy detector suggests that solar wind electrons very likely also dominate the >100 keV energy channels.

Although this letter focuses on presenting new observational evidence for quasi-parallel shock-acceleration of electrons, here we point out selected features that are relevant for our understanding of the acceleration process. The low-energy electron energisation near the shock front was associated with high-energy electrons detected in a broader region that extends both upstream and downstream (Fig. 2d). Short, Large-Amplitude Magnetic Structures (SLAMS)[22] were observed in this same region (Fig. 2a), raising the possibility that the interaction between high-energy electrons and these magnetic structures may have played a role in the acceleration to the highest measured energies. The change in the shape of the shock-accelerated electron spectrum at ~100 keV (Fig 4.) suggests that electrons at this energy and higher were potentially subject to a different acceleration process (note that the weak "background" signal of escaping magnetospheric electrons does not provide an obvious explanation for this). For example, electrons at this energy and above may have had gyroradii large enough to be efficiently scattered by SLAMS. We note that the high-energy electron profile of this shock encounter is not in agreement with diffusive shock acceleration theory, which predicts a constant downstream flux at any given energy. One of a number of possible explanations for this is that these observations were not made under steady-state conditions.

Acknowledgments: We thank the relevant Cassini instrument Principal Investigators (not included in the author list) who are responsible for the data sets used: D. A. Gurnett, S. M. Krimigis, and D. T. Young. This work was supported by UK STFC through rolling grants to MSSL/UCL and Imperial College London.

Author contributions: A.M. identified the event, analysed the combined data set, proposed the interpretation, and wrote the paper. L.S., M.F., S.J.S., and H.H. discussed the interpretation. N.S., M.F.T., A.R., and G.R.L. each analysed, and checked the interpretation of, a single data set. A.J.C., P.C., and M.K.D. oversaw the data interpretation. All authors discussed the results and reviewed the manuscript.

Additional information: The data reported in this paper are publicly available in NASA's Planetary Data System at http://pds.nasa.gov. Correspondence should be addressed to A.M.

Competing financial interests: The authors declare no competing financial interests.

**Figure 1. Overview of the spacecraft encounter with Saturn's bow shock on 3 February 2007 (not to scale).** In Kronocentric Solar Magnetospheric (KSM) coordinates (origin at the centre of the planet, *x*-axis pointing towards the Sun, *z*-axis chosen to define an *xz* plane containing Saturn's magnetic dipole axis, and *y*-axis completing the right-handed Cartesian set) the spacecraft location at the time of the shock crossing was $(x, y, z) \sim (17, 4, -3)$, in units of Saturn radii ($R_S$). In the zoom-in view of the shock encounter the increasing grey shading from left to right indicates the increase in thermal plasma density across the shock, and the simplest possible spacecraft trajectory in the shock rest frame is shown.

**Figure 2. Observations made during the shock crossing (00:00 to 02:00 UT). a,** Magnetic field components (KSM coordinates)[23]. Between 00:00 and 00:05 UT the mean magnetic field (in units of nT) was $(B_x, B_y, B_z) \sim (0.03, 0.05, -0.07)$, giving $\theta_{Bn} \sim 60°$, and between 00:05 and 01:00 UT it was $\sim (0.1, 0.02, -0.05)$, giving $\theta_{Bn} \sim 20°$ ($\theta_{Bn}$ based on a model-predicted shock normal[17]). **b,** Magnetic field magnitude. **c,** Frequency-time spectrogram of electric field Power Spectral Density (PSD)[24]. The electron gyrofrequency is over-plotted in black, and estimates of the upstream electron plasma frequency and measurements of the downstream electron plasma frequency are over-plotted as dashed white lines. **d** and **e,** Energy-time spectrograms of electron Differential Intensity (DI) for high and low-energy

ranges, from the Low-Energy Magnetospheric Measurements System (LEMMS)[25] and anode 5 of the Cassini Electron Spectrometer (ELS)[27], respectively. The ELS energy range has been restricted to <18 keV. Vertical dashed lines indicate sub-intervals relevant for Fig. 3. **f**, Normalized electron DI in three energy ranges. In the 5-10 keV range only data above the ELS one-count level is shown. **g**, Energy-time spectrogram of ion Differential Energy Flux (DEF)[26].

**Figure 3. Average electron spectra for different time intervals during the crossing.** In all panels the vertical dashed line separates data from the low-energy (ELS) and high-energy (LEMMS) instruments. Dashed curves give the ELS one-count level and dotted curves show Maxwellian distributions for comparison. In **a** the single peak at low energy is a mixture of solar wind electrons and spacecraft photoelectrons (that result from photons hitting the metallic surfaces of the spacecraft). In **b** and **c** the two peaks at low-energy indicate separation of these populations (where the spacecraft photoelectron population is less energetic). Grey rectangles give the energy ranges of the LEMMS energy channels. LEMMS data have been background-subtracted, and no data point in an energy channel indicates intensity at the background level. The lack of inter-calibration between the instruments produces an offset in differential intensity at 18 keV, which we do not attempt to address because ELS was not significantly above the one-count level at 18 keV at any point during the encounter. Error bars are standard deviations for each sub-interval.

**Figure 4. Comparison of high-energy electron spectra.**

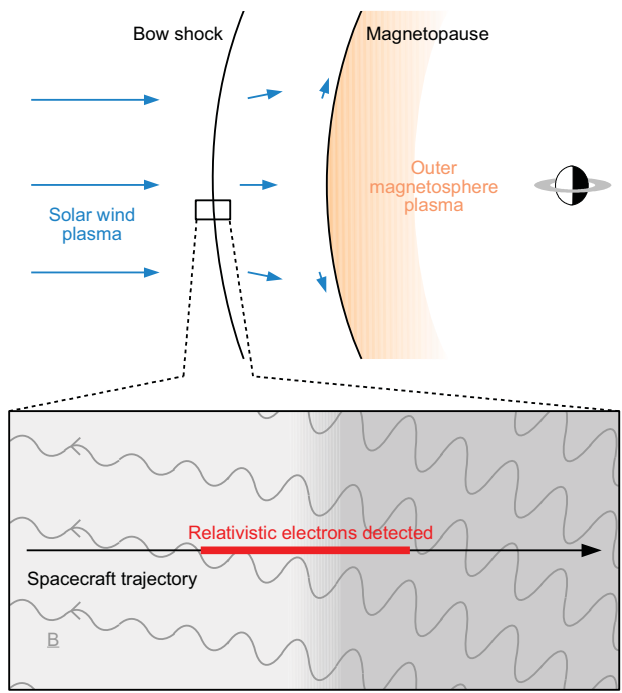

Bow shock

Magnetopause

Solar wind
plasma

Outer
magnetosphere
plasma

Relativistic electrons detected

Spacecraft trajectory

B

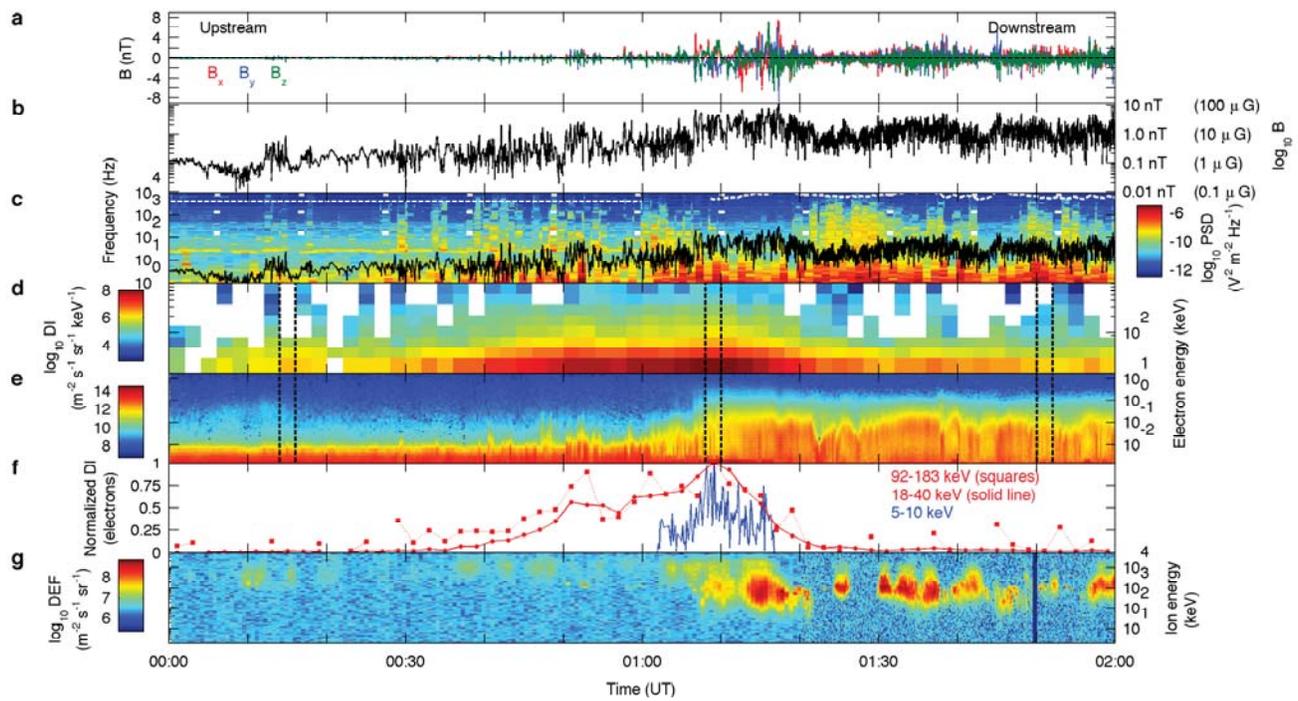

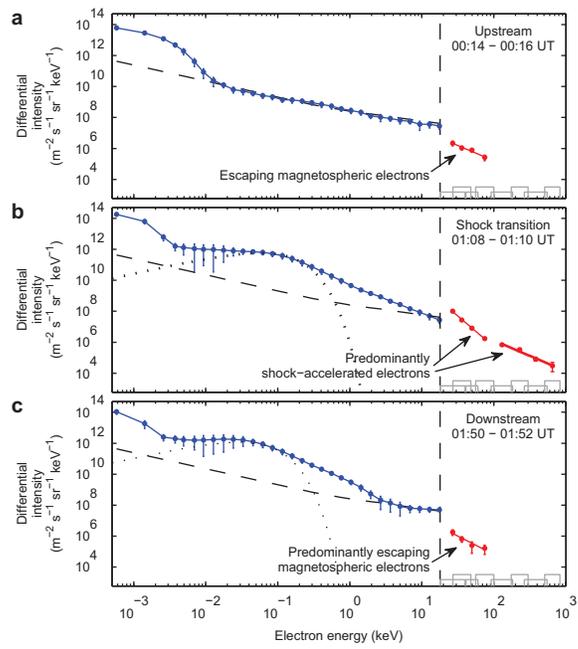

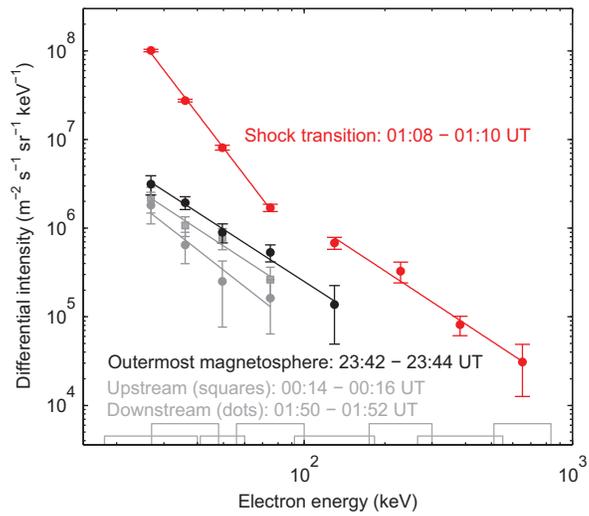